# Clock Auto-synchronizing Method for BES III ETOF Upgrade*

WANG Si-Yu(王思宇)[1,2]　　CAO Ping(曹平)[1,2;1]　　LIU Shu-Bin(刘树彬)[1,2]　　AN Qi(安琪)[1,2]

[1] State Key Laboratory of Particle Detection and Electronics
[2] Anhui Key Laboratory of Physical Electronics, Department of Modern Physics, University of Science and Technology of China, HeFei 230026, China

**Abstract:** An automatic clock synchronizing method implemented in field programmable gate array (FPGA) is proposed in this paper. It is developed for the clock system which will be applied in the end-cap time of flight (ETOF) upgrade of the Beijing Spectrometer (BESIII). In this design, an FPGA is used to automatically monitor the synchronization circuit and deal with signals coming from external clock synchronization circuit. By testing different delay time of the detection signal and analyzing state signals returned, the synchronization windows will be found automatically in FPGA. The new clock system not only retains low clock jitter which is less than 20ps root mean square (RMS), but also demonstrates automatic synchronization to the beam bunches. So far, the clock auto-synchronizing function has been working successfully under a series of tests. It will greatly simplify the system initialization and maintenance in the future.
**Keywords:** BESIII, endcap time-of-flight, clock synchronization
**PACS:** 29.85.Ca

## 1 Introduction

The Beijing Electron Positron Collider (BEPC) and the Beijing Spectrometer (BES) [1], [2] are upgraded to BEPCⅡ and BESⅢ [3]-[5] respectively since the summer of 2008. The time-of-flight (TOF) system, with the physical goal of particle identification (PID), is a very important part of the BESⅢ. To improve the time resolution of PID, the newly developed gaseous and widely used detector, multi-gap resistive plate chamber (MRPC) [6-8], is chosen for the upgrade ETOF detector instead of plastic scintillator bars read out by fast fine mesh photomultiplier tubes (PMT). After upgrade, the total time resolution will be improved significantly from 138ps in total to better than 80ps, and among which only 25ps is limited to be caused by electronics [9].

To ensure the 25ps time resolution of the TOF electronics, the clock jitter must be less than 20ps RMS and the clock phase should be highly synchronized to the beam collision time [10]. As MRPC detectors are utilized in upgrade system, there is a significant increase in the number of electronic channels and two more VME64xP crates as well as another two clock modules will be needed to be dedicated for ETOF electronic system [11]. Thus, the original clock system will be unable to meet the needs of the upgraded system other than upgrading it with more clock output channels. On the other hand, each time the original TOF system is powered on, it requires a series of manual operations to configure the clock synchronization which is not very intelligent. As the TOF clock system is under upgrade, the configuration operations will be simplified with the help of a new algorithm implemented in FPGA.

## 2 Automatic clock synchronizing method

### 2.1 Proposed Clock System

To meet the needs of TOF electronics, the achieved clock system consists of two parts - one is the transmission of the RF signal, and the other is VME clock modules which are responsible for providing multi-channel high quality clocks as well as synchronization and phase-control among them.

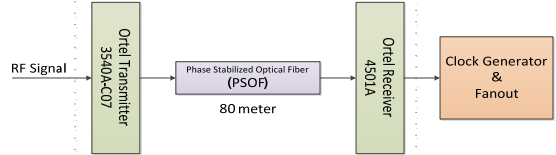

Fig. 1 Block diagram of TOF clock system

Fig.1 illustrates the block diagram of the whole TOF clock system. The accelerator provides RF clock signal which is transmitted by optical fiber. To minimize the effects of temperature change, the optical transmitters and receivers from Ortel company and the Phase-Stabilized Optical Fiber (PSOF) from Furukawa company have been utilized. At last, the RF clock signal is transmitted to VME clock modules for clock generation and fan-out.

Considering the structure of the upgraded TOF read-out system, the new system will be made up of four clock modules-one is master module and the others are slave modules. In addition, the master module is also required to be able to generate clocks and make them synchronized for the whole

* Supported by National Natural Science Foundation of China (10979003, 11005107)
1) E-mail: cping@ustc.edu.cn

TOF read-out electronic system. To simplify design, both master and slave clock modules share the same circuit design, of which the scheme is shown in Fig.2.

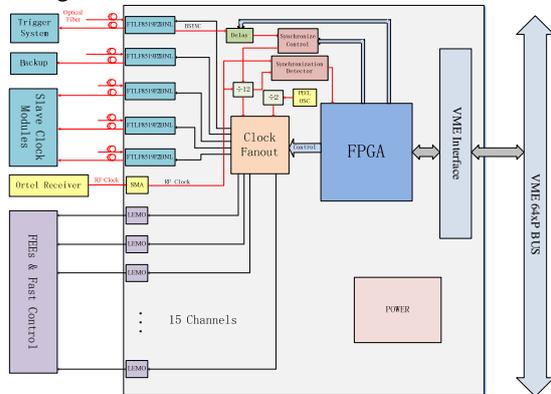

Fig. 2 Scheme of the clock module

By selecting different clock sources of the clock fan-out chip, clock module will work under master or slave mode. In master mode, system clock is the 499.8MHz RF clock divided by 12 from the accelerator; in slave mode, it is a 41.67MHz optical signal output from the master clock module. Besides, every clock module has an 83.3MHz crystal oscillator onboard for clock generation by itself under the off-line mode. The clock fan-out chip SY89829 has 20 channels output, of which five channels are transformed to optical signals for trigger systems and slave modules, and the other fifteen channels are transmitted in the form of LVPECL electrical signals to other VME modules which are located in the same VME64xP crate.

An FPGA is also used for system control. It supports communication between computer and electronic system via VME interface protocol.

### 2.2 Clock Synchronizing and Monitoring

The period of beam bunches in accelerator is 8ns and the synchronized RF signal is 499.8MHz. As mentioned before, TOF clock is generated by a simple divider so that there are four possible phases between the beam bunches and the TOF clock. To get a constant phase of TOF clock, a BSYNC signal is derived from accelerator for phase adjustment of which leading edge contains the phase information of beam bunches. Therefore, clock synchronization will be achieved once the time interval between the leading edge of BSYNC and that of the TOF clock is determined. The simplified diagram of synchronization control and clock generation circuit is designed as Fig.3 shown.

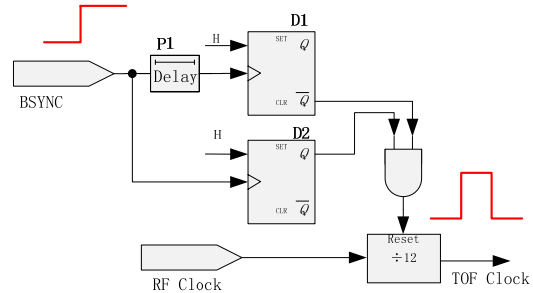

Fig. 3 Simplified diagram of synchronization control and clock generation circuit

A periodic clock signal is formed by two D-flip-flops (DFFs) which convert the TOF clock duty from 1/2 to 1/12. This new signal has the same phase with the TOF clock so that it could be used to check current clock synchronization state, as shown in Fig.4.

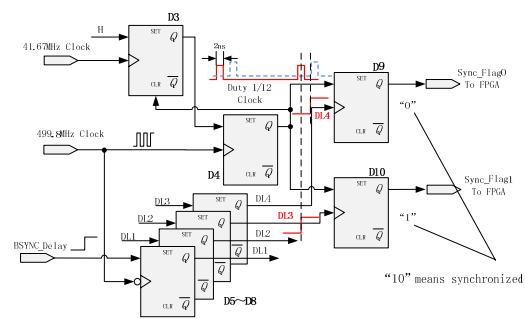

Fig. 4 Simplified diagram of synchronization monitor circuit

Once the clock phase is determined, we can get a synchronization window by adjusting the delay value of the BSYNC signal. Theoretically the window width should be 2ns, but considering the instability of the edge of signals output from the DFFs, the actual window width will be slightly less than 2ns. Corresponding to the different phases of TOF clock, there are more than one synchronization windows and all of them are connected one by one on the delay-time axis. However, the final synchronization window that we choose is determined by the phase of the TOF clock which we desired.

### 2.3 Algorithm in FPGA

In the previous TOF clock system, a lot of VME read and write operations need to be done manually for measuring synchronization windows, as well as the calculation of the center value. Even a small change in system, for instance, a replacement of transmission cables, will require the synchronization windows to be re-measured. To simplify the operation of synchronization, a method of automatic clock synchronizing implemented in FPGA has been processed.

As mentioned above, the BSYNC signal is delayed by a SY89295 chip. The chip is a programmable delay line that delays the input signal using a 10-bits-long digital control signal. Then, the synchronization state will be adjustable by changing different configuration data from FPGA to the delay chip. The automatic synchronization starts with a reset signal of dividers according to an initial delay data. If the feedback synchronization flag (SynFlag) is '10', then the rest operations can be continued; otherwise, the initial delay data should be a slight increase in value to make the system working in a steady state of some specific synchronization window.

Fig.5 shows a flowchart for automatic synchronization logic which is mainly consist of two parts - Step 1 and Step 2. The purposes of them are to measure the maximum and the minimum values of synchronization window respectively. By changing the delay value from coarse count to fine count and testing whether the value of SynFlag is '10', however, the boundary values of the synchronization interval will be found. Finally, the center value can be easily calculated by averaging the boundary values, then can be used as the delay value for synchronization calibration that make sure the whole TOF system working under the same clock phase after powered on.

On both sides of the boundary of the synchronization window, there is a short interval unstable and the state of which cannot be determined whether it is actually in the range of synchronization window just by a single detection. Thus during measurement of the maximum and minimum values, the result will be considered correct only if the synchronization flag we got is "10" for more than 8 times. Figure 6 shows a schematic of TOF clock synchronization window.

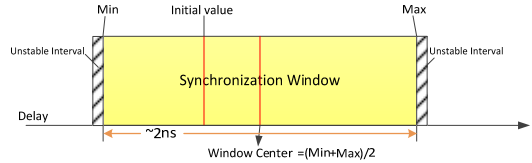

Fig. 6 Schematic of synchronization window

This procedure can be called for controlling the TOF clock synchronization every time after system power-on. With the power keeping supplying, we can also call this logic module to re-measure the synchronization window automatically by VME reading and writing. Meanwhile, the new logic can not only test the representative values of the synchronization window automatically, but also remained the functions of manual operation and detection as the old TOF clock system does, which means that we can get the information in both ways and verify if the results we got are reliable.

## 3 Test Results

For the new TOF clock module testing, two optical fibers in different length are used for transmitting the Pickup signal in previous TOF system. The synchronization information is shown as Table 1. There are significant differences in the positions of two synchronization windows which are caused by the different initial phases between pickup signal and RF clock. 1 bit in delay chip corresponds to 9ps approximately so that the actual delay values in table are obtained by data in decimal number multiplied by 9ps.

Table 1 Synchronization of different optical fibers

| Register | Function | Delay Using Fiber A | | Delay Using Fiber B | |
|---|---|---|---|---|---|
| | | (0xX) | (ps) | (0xX) | (ps) |
| 0xf050 | Center | 0x174 | 3348 | 0x11d | 2565 |
| 0xf0c0 | Minimum | 0x108 | 2376 | 0xb0 | 1584 |
| 0xf0d0 | Maximum | 0x1e0 | 4320 | 0x18b | 3555 |

To get all synchronization windows, different initial delay values are provided. As mentioned before, there are unstable intervals in the edge and that may cause overlaps of adjacent windows. After improving the correction algorithm by measuring repeatedly near the edge of synchronization windows, the overlaps are all gone, as shown in Table 2.

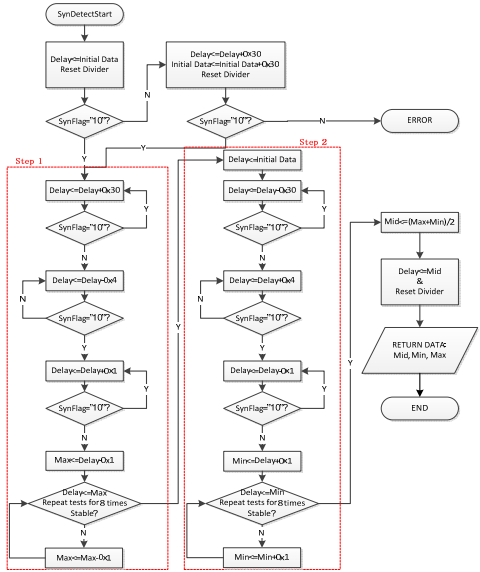

Fig. 5 Logic flowchart of automatic synchronization

Table 2 Detection of synchronization windows

| Register | Function | Before correction | | | After correction | | |
|---|---|---|---|---|---|---|---|
| | | Window1 | Window2 | Window3 | Window1 | Window2 | Window3 |
| 0xf040 | Initial | 0x100 | 0x1c0 | 0x290 | 0x100 | 0x1a0 | 0x280 |
| 0xf050 | Center | 0x11d | 0x1fc | 0x2d5 | 0x10d | 0x1e3 | 0x2b9 |
| 0xf0c0 | Minimum | 0xb0 | 0x18c | 0x268 | 0xa4 | 0x178 | 0x250 |
| 0xf0d0 | Maximum | 0x18b | 0x26c | 0x343 | 0x177 | 0x24e | 0x322 |

When the synchronization configuration is done, the rising edge of the delayed pickup signal will decide the system clock phase. The specific waveform of pickup_delay signal and 1/12 40MHz Clock signal is illustrated in Fig.7. The system clock has been demonstrated to be well synchronized to the beam bunches.

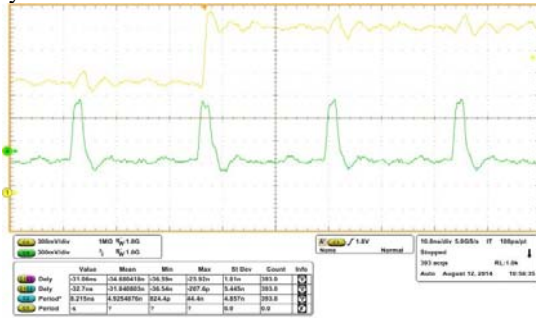

Fig. 7 Waveform of synchronized clock and pickup_delay

Since there is almost no change in the clock module circuit, the system clock jitter has retained to be less than 20ps RMS. It has been already reconfirmed by new tests.

## 4 Conclusions

In this paper, a method of automatic clock synchronizing implemented in FPGA is proposed for clock system during BESIII ETOF upgrade. It combines both FPGA algorithm and external off-the shelf devices. The synchronization intervals can be calculated automatically by adjusting external delay chip and analyzing the returned synchronization flags. According to test results, the function of clock automatic synchronizing to the beam collision time has been achieved without any influences to the system clock quality.

*The authors gratefully acknowledge all of the BES III collaborators who helped to make this work possible.*

## References

[1] J. Z. Bai et al. Nucl. Instrum. Methods A, 1994, 344(2): 319–334
[2] J. Z. Bai et al. Nucl. Instrum. Methods A, 2001, 458(3): 627–637
[3] F. A. Harris, BES Collaboration, "BEPCII and BESIII," Nucl. Phys.s B (Proc. Suppl.), vol. 162, pp. 345–350
[4] W. Li, BES Collaboration, in Proc. 4th Flavor Phys. CP Violation Conf. (FPCP 2006), Vancouver, BC, Canada, 2006
[5] BES Collaboration, "The construction of the BESIII experiment," Nucl. Instrum. Methods A, vol. 598, no. 1, pp. 7–11, Jan. 2009
[6] Williams M C S, Cerron E et al. Nucl. Instrum. Methods A, 1999, 434(2–3): 362
[7] CHEN Hong-Fang, LI Cheng et al. High Energy Physics and Nuclear Physics, 2002, 26(3): 201 (in Chinese)
[8] WU Jian, Bonner B et al. Nucl. Instrum. Methods A, 2005, 538: 243
[9] Huanhuan Fan et al. IEEE Trans. Nucl. Sci., 2013, PP(60): 1-7
[10] Hao Li et al. IEEE Trans. Nucl. Sci., 2010, 57(2): 442-445
[11] P.Cao et al. Chinese Physics C., 2014 38(4), pp. 046101